# Optical Properties of an IR-Transparent Prismatic Plate in Exposure to the Thermal Radiation and its Application as a Radiation Rectifier

A. Raghavi

Department of Physics, Payame Noor University, Tehran 19395-4697, Iran

Abstract

Optical properties of a prismatic transparent plate and its effects on thermal radiation heat transfer are studied. A 3D ray-tracing numerical method is used to prove the bi-transmittance property of the prismatic sheet when it is exposed to thermal radiation in two opposite directions. It is shown that in the present of a prismatic window a thermal equilibrium with a temperature gradient between two sources is possible.

## I. Introduction

Metamaterials are artificial structures (products of human ingenuity), designed to obtain controllable electromagnetic or optical properties. It includes designing and constructing artificial media with properties not found naturally [1]. The main index of this kind of materials is their abnormal behavior which is governed by their structure rather than their composition. According to this generalized definition, the new optical components known as prismatic structures can also be classified as a type of metamaterials, designed artificially for specified applications in light control. A prismatic sheet is an array of prismatic elements that are positioned next to each other in a 2D or 3D pattern [2]. Prismatic sheets can be categorized in two distinct subgroups as microprismatic and macroprismatic structures, referring to the size of individual prisms relative to the wavelength of the applied radiation. Microprismatic sheets differ from the other one in both the application domain and the method which is applicable for their studies. The methods which are useable in microprismatic structure analyzes usually are based on electromagnetic theory and wave optics [3], while for analyzing the properties of macroprismatic devices the methods of geometric optics are sufficiently accurate and acceptable [4].

In the present work a different potential application for the transparent macroprismatic sheets is introduced and analyzed which can work as a thermal radiation rectifier. It is shown by means of an extended numerical simulation approach that a prismatic transparent sheet has an abnormal bi-transmittance property, i.e. different transmittance coefficients in two opposite directions, against the natural thermal radiation of bodies.

The paper is organized as follows. In section II bi-transmittance sheets and their effects on thermal equilibrium state between two radiation sources is explained. Section III is devoted to introducing and analyzing the transmission properties of a prismatic sheet, using a 3-D ray-tracing matrix method. More realistic conditions for radiation and energy exchange between two thermal sources are studied in section IV, where the parameters such as emissivity of the surfaces, spectral content of radiation and the material of prismatic plate are bring into account. A complementary argument about the time scale of equilibrium process and effects of other competitor phenomena in heat transfer mechanism is accessible in section V. Finally, a summary and conclusion is presented in section VI.

## II. Thermal equilibrium in presence of a bi-transmittance medium

Before starting the main work which is the abnormal optical properties of the prismatic sheets, let us drop some lines about the feature of such an abnormal medium in accordance with the radiation heat transfer process. This preliminary argument is worthwhile because the foremost use of the under-study metamaterial is realized to be placed in this field of applications.

Two infinitely long and directly opposed parallel surfaces at temperatures $T_i$ are in thermal energy exchange with a hemispherical total radiant obeying the Stefan–Boltzmann law,

$$E_i(T) = \varepsilon_i \sigma T_i^4 . \tag{1}$$

Here, $\sigma$ is the Stefan–Boltzmann constant and $\varepsilon_i$ is the total hemispherical emissivity for i'th surface. If two sources initially be in different temperatures then different energy flux between two surfaces bring the total system to a final equilibrium state with both surfaces in the same equilibrium temperature. This is the ordinary thermal radiation equilibrium process which occurs in a normal thermal radiation transfer mechanism. If an ordinary transparent sheet is placed between two sources, as it is shown schematically in Fig. (1), then still the state of the equilibrium will be the same as before, unless the relaxation time which will increase because of the decreasing the transferring energy.

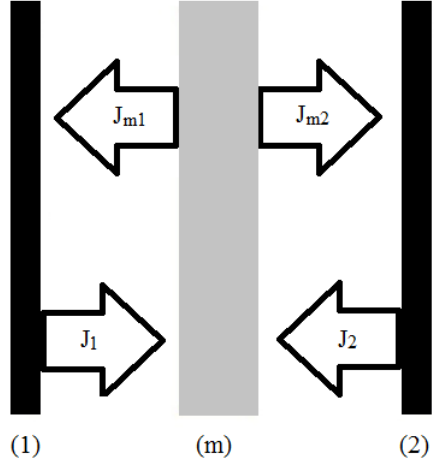

Figure 1: General arrangement of a radiation heat transfer system in the present of a semitransparent medium. Surfaces (1) and (2) are radiating sources and the sheet (m) is the semitransparent medium.

From the other hand, if the transparent sheet in the previous structure supposed to be an unusual medium with two different total hemispherical transmittances $\tau_{12}$ and $\tau_{21}$ for two opposite directions then the process of energy transfer between two surfaces will be altered with a quite different consequence. To find the equilibrium temperature for new arrangement of the system a net-radiation method [5] is employed to analyze the radiant heat transfer between two sources. The surfaces (1) and (2) in Fig. 1 are supposed to have the same emissivity $\varepsilon$ and reflection index $\rho = 1 - \varepsilon$ and are in thermal exchange only through their thermal radiation. The medium (m) is a bi-transmittance window with the absorption coefficient $\alpha_m$ and the reflectivity indexes $\rho_{m1} = 1 - \alpha_m - \tau_{12}$ and $\rho_{m2} = 1 - \alpha_m - \tau_{21}$.

Introducing the total emerging radiation power per unit area for surfaces (1) and (2) as $J_1$ and $J_2$ and for two faces of (m) as $J_{m1}$ and $J_{m2}$ then from the conservation of energy the following equations hold,

$$J_1 = \varepsilon E_1 + \rho J_{m1} \qquad (2)$$
$$J_2 = \varepsilon E_2 + \rho J_{m2} \qquad (3)$$
$$J_{m1} = \varepsilon_m E_m + \rho_{m1} J_1 + \tau_{21} J_2 \qquad (4)$$
$$J_{m2} = \varepsilon_m E_m + \rho_{m2} J_2 + \tau_{12} J_1 \qquad (5)$$

Here, $E_1$, $E_2$ and $E_m$ are the thermal radiant of surfaces (1), (2) and (m), respectively. The temperature of both sides of (m) is taken to be the same. Because both diffusive and specular reflection and transmission are present, the view factors for the faces are too difficult to be included. To avoid this difficulty we set all view factors equal to unity by extending all surfaces infinitely. It is equivalent in practice to place whole of the system inside an enclosure with fully reflective walls which resembles the problem to the same situation as the case of infinitely extended surfaces. The net thermal radiation flux, entering the surfaces (1), (2) and (m) then are

$$q_1 = \alpha J_{m1} - \varepsilon E_1 \qquad (6)$$
$$q_2 = \alpha J_{m2} - \varepsilon E_2 \qquad (7)$$
$$q_m = \alpha_m(J_1 + J_2) - 2\varepsilon_m E_m, \qquad (8)$$

where from the Kirchhoff's law, $\varepsilon = \alpha$ and $\varepsilon_m = \alpha_m$. When whole of the system is in thermal equilibrium the net energy exchange between surfaces are zero i.e.

$$E_1 - J_{m1} = 0 \qquad (8)$$
$$E_2 - J_{m2} = 0 \qquad (9)$$
$$2E_m - (J_1 + J_2) = 0. \qquad (10)$$

Solving the recent set of equations together with equations (2-5) we get to the following relation between the radiant of two sources,

$$E_2 = \frac{\varepsilon_m + 2\tau_{12}}{\varepsilon_m + 2\tau_{21}} E_1 \qquad (11)$$

or by substituting from Eq. (1),

$$T_{e2} = \left(\frac{\varepsilon_m + 2\tau_{12}}{\varepsilon_m + 2\tau_{21}}\right)^{1/4} T_{e1}. \qquad (12)$$

Here $T_{e1}$ and $T_{e2}$ are the equilibrium temperatures of surfaces (1) and (2), respectively. For the case of an ordinary medium with $\tau_{12} = \tau_{21}$, this leads to the expected result, $T_{e1} = T_{e2}$. But the interesting result achieves when the medium is taken to be a bi-transmittance medium with $\tau_{12} \neq \tau_{21}$. In this situation Eq. (12) says that the thermal equilibrium occurs for even though two sources are in different temperatures. In what follows it is tried to prove that a transparent prismatic sheet can be a practically available exemplar for a real bi-transmittance medium.

### III. Optical properties of a prismatic sheet

In this section we are going to prove the bi-transmittance property of transparent prismatic planes. An adhoc 3D ray-tracing matrix code is developed for this purpose along the lines of the formulation that is introduced in [4]. The code is used to investigate the behavior of transmittance of a prismatic transparent sheet when it is exposed to a flux of thermal radiation with a Lambertian angular distribution.

An illustrative picture of what we mean from a prismatic sheet is depicted in Fig. (2). The corrugations on surface of plate are pyramidal prisms with the same apex angles. The apex angle of the prisms is an important parameter that governs its optical properties. A prismatic sheet with apex angles of $180°$ is a flat parallelepiped window with the same transmittance for two oposite directions.

Nevertheless, for other apex angles the equality or inequality of two transmittances remains as a question to be answered by more investigations.

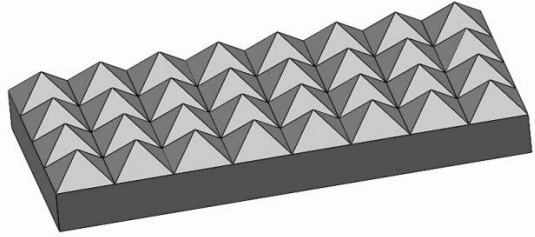

FIG. 2. A prismatic plate.

It is important to be mentioned here that the parameter that is calculated in this manner as transmittance of a medium is not exactly the same as the transmittance of radiation energy. Strictly speaking, what the ray-tracing method gives is just the transmittance of rays which can be regarded as a representation for the transmittance of photonic flux. Relation of two kinds of transmittances is discussed in more details in section III.

The real sample that is used in the present study is an infinitely extended plate which can be achieved in simulation by placing a unit prismatic cell inside a waveguide with totally reflecting walls, as it is illustrated in Fig. 3.

In practice, the effects of boundaries should be avoided by either increasing the reflectivity of walls or increasing the total dimensions of the sample prismatic sheet compared to the dimensions of a unit cell.

The arrangement of different components in the system which is employed to evaluate the transmittance of the prismatic plate is sketched in Fig. (3). The radiation sources (a) and (b) are two plane surfaces with the same physical properties and parallel to the base of the prism. The images of sources and unit prism in the enclosure walls simulate the situation of two un-bounded radiating surfaces and an infinitely extended prismatic plate.

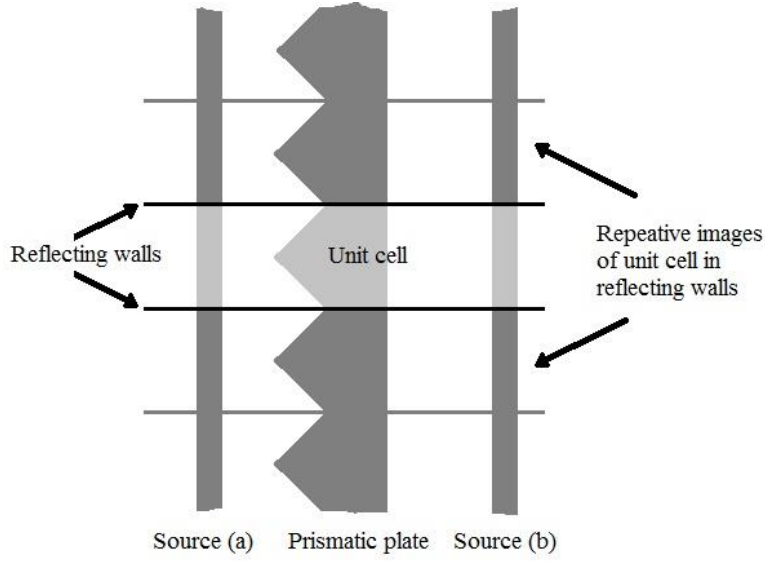

FIG. 3. Schematic arrangement of the under study system. Surfaces (a) and (b) are radiating sources and prismatic plate is the transparent medium.

Radiation is simulated in the form of emerging rays originating form randomly chosen points in the surface of radiation source toward the other surface. In this stage we do not include the spectral content of the radiation, since it is discussed in details in the next section.

However, the angular distribution of radiation is an important parameter that must bring into account carefully. The angular distribution of the thermal radiation is known to be Lambertian, with an angular radiation intensity obeying the Lambert cosine law [5]

$I(\theta) = I_0 \cos(\theta)$ (13)

where $\theta$ is the polar angle and $I_0$ is the intensity in normal direction. To achieve the desired angular distribution for generated rays first both polar and azimuthal angles $(\theta, \varphi)$ are generated randomly and uniformly in the ranges of $[0, \frac{\pi}{2}]$ and $[0, 2\pi]$, respectively. Then the distribution of polar angles $\theta$ is reformed to the cosine distribution (13) using the method of inverse-transform with the corresponding transformation function, $\theta \to \sin^{-1}\theta$ [6]. The resultant will be a randomly uniform distribution of rays which are emerging from the surface in random directions with cosine distribution in polar angles and uniform distribution in azimuthal angles.

Since the thermal radiation in the room temperature occurs in the range of IR spectrum, the constituting material of prism are chosen from an IR transparent material for a room temperature application. More in vogue materials for this purpose are IR transparent crystals with a refraction index about $\approx 1.5$, (e.g. NaCl and KBr) and $n \approx 2.4$ (e.g. ZnSe and KRS5) or IR transmitting glasses such as arsenic glasses ($n \approx 2.5$). Accordingly we choose to work with two sample values, $n = 1.5$ and $n = 2.4$ for the refraction indexes to show the effect of material on desired properties.

The reflectivity for an un-polarized directional wave from each interface obtains from the Fresnel's relations

$$\rho = \frac{1}{2}\left[\left(\frac{n^2 \cos\theta - \sqrt{n^2 - \sin^2\theta}}{n^2 \cos\theta + \sqrt{n^2 - \sin^2\theta}}\right)^2 + \left(\frac{\cos\theta - \sqrt{n^2 - \sin^2\theta}}{\cos\theta + \sqrt{n^2 - \sin^2\theta}}\right)^2\right] \quad (14)$$

where, $n = \frac{n_2}{n_1}$ is the relative refraction index of the interface [5]. The Fresnel's reflection from the interfaces of two mediums is included in the code by assigning a reflection probability amplitude to each ray, equal to the reflectivity which obtains from (14).

The absorption inside the prism is neglected for the sake of simplicity ($\varepsilon_m = 0$). Of course it wouldn't have a significant effect on our results because practically it is negligible for a good transparent medium. Moreover, the contribution of both sides of plate to the net radiation received by two sources are the same, regardless the shape of its surfaces.

In the numerical 3D ray-tracing method each ray is specified with three components of its unit vector (direction cosines) together with the start and end points on the surfaces that constitute the physical structure of the system. Since all surfaces are flat, they are totally determined from the surface equation in Cartesian coordinate as

$$N_x x + N_y y + N_z z + D = 0 \quad (15)$$

where the component of the normal unit vector, i.e. $N_x$, $N_y$, $N_z$ together with D specify both the orientation and the position of each surface, exactly. The line path of the ray should be also presented with the general line equation

$$\vec{r} = \vec{r}_\circ + t\hat{d} \quad (16)$$

where $\vec{r}_\circ$ is the start point, $\hat{d}$ is a unit vector indicating the direction of ray propagation and t is the line parameter.

The start point for each ray is a randomly chosen point on one of the sources and its direction is a randomly direction with a cosine distribution in polar angle, as it was discussed previously. To find the end point we let the line equation of ray (Eq. 16) to intersect with the plane equations of all surfaces (Eq. 15) and then choose the nearest intersection point as the end point of the ray. In the next step we set the end point of the old ray as the start point of a new ray and try to find the new direction according to the fact that it is whether reflected from the interface or refracted into the new medium. The new direction obtains simply by using the laws of reflection and refraction (Snell's law) respectively. In practice, for a set of incident, reflected and refracted rays which can be indicated by unit vectors $\hat{r}_i(i_x, i_y, i_z)$, $\hat{r}_r(r_x, r_y, r_z)$ and $\hat{r}_t(t_x, t_y, t_z)$, respectively, we can use the matrix equations

$$\begin{pmatrix} i_y N_z - i_z N_y & -(i_x N_z - i_z N_x) & i_x N_y - i_y N_x \\ N_x & N_y & N_z \\ i_x & i_y & i_z \end{pmatrix} \times \begin{pmatrix} r_x \\ r_y \\ r_z \end{pmatrix} = \begin{pmatrix} 0 \\ \cos \alpha_r \\ \cos \delta_{ir} \end{pmatrix} \quad (17)$$

and

$$\begin{pmatrix} i_y N_z - i_z N_y & -(i_x N_z - i_z N_x) & i_x N_y - i_y N_x \\ N_x & N_y & N_z \\ i_x & i_y & i_z \end{pmatrix} \times \begin{pmatrix} t_x \\ t_y \\ t_z \end{pmatrix} = \begin{pmatrix} 0 \\ \cos \alpha_t \\ \cos \delta_{it} \end{pmatrix} \quad (18)$$

where $\cos \alpha_r = \hat{r}_r \cdot \hat{N}$, $\cos \alpha_{ir} = \hat{r}_i \cdot \hat{r}_r$, $\cos \alpha_t = \hat{r}_t \cdot \hat{N}$ and $\cos \alpha_{it} = \hat{r}_i \cdot \hat{r}_t$ [4]. The square matrix in Eqs. (17) and (18) is composed of characteristics of the normal incident ray, plane of incidence, and incident ray. Referring to this as the incident matrix $M_{in}$ then the reflection and the refraction vectors M can be derived from the formal solution to the matrix equations (17) and (18) as

$$M = M_{in}^{-1} M_p \quad (19)$$

where, $M_p$ is either of angle vectors to the right of Eqs. (17) and (18).

When the new direction is determined through Eq. (19), we can repeat the previous steps, until either the ray passes through the prism and hit the next source in the opposite side or reflects to the first side and return back to the source. In the former case we count the ray as a transmitted ray while in the latter case it will reckoned as a reflected ray.

Sometimes, but very seldom, it happens for a ray to fall in an endless or very long loop of reflections and refractions without ending to any one of source plates. To avoid this situation we set a finite number (e.g. nt=100) as the allowed number of distinct propagation of rays.

The above mentioned algorithm which is the base of our code is exploited to solve the problem of radiation transmission through a prismatic plate in a geometry as shown schematically in Fig. (3).

Before starting the main simulation study lets to check the degree of accuracy of the code result by comparing it with a known theoretical value, namely the reflectance of a plane transparent parallelepiped window. For a non-absorbing transparent plane window with directional reflectivity $\rho$, the total directional transmittance reads [5]

$$T = \frac{1-\rho}{1+\rho} \quad (20)$$

For example when n=2.4 and the incident radiation is $\theta = 30°$ equation (14) gives $\rho = 0.17$. Substituting this into (20) one find T=0.7079 for the total transmittance.

Now we choose the same refraction index and incident angle as the input parameters of our code and then execute it for an apex angle of $\alpha = 180°$ (flat sheet). The result is T=0.7107 with a good agreement with the corresponding theoretical value. Repeating the test for other samples give the same convincible results.

After getting assurance about the reliability of the code now we can start our main study. The number of rays for each execution is chosen to be $N = 10^5$, for which a stable result obtains that do not varies notably by increasing the ray number $N$. Simulation is executed for different apex angles from $\alpha = 60°$ up to $\alpha = 180°$ with a length step of $5°$. The simulation is repeated for two sample refraction indices n=1.5 and n=2.4 separately. The results are graphed in Figs. (4a&b), where total transmittance is plotted versus the apex angle of prisms for both 'apex to base' (dot-dashed line) and 'base to apex' (dashed line) directions. The difference between two transmittance values is also plotted in a solid line. Error, which is the number of truncated rays because of lengthening their paths, is indicated by dots for each apex angle.

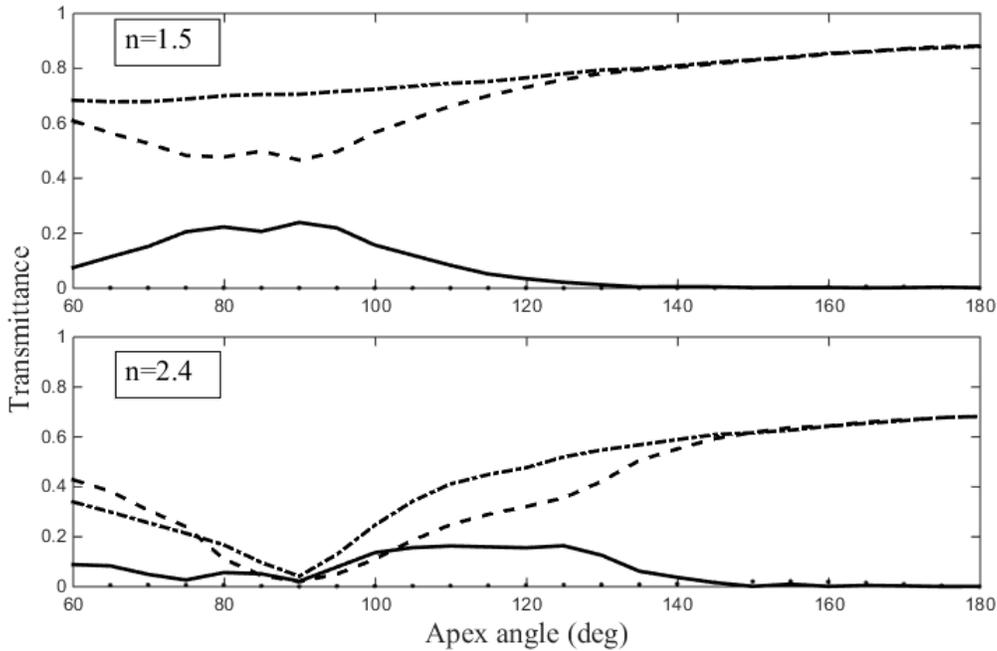

FIG. 4. Ray transmittance of a prismatic sheet in two opposite directions for two different refraction indices.

As it is evident from these illustrations, a transparent prismatic plate is really a bi-transmittance medium. Solid lines in two figures show that the transmittance difference is a function of both the refraction index and the apex angle of prisms. Maximum difference occurs in different apex angles for different constituting materials. For n=1.5 the prism with maximum difference in two transmittances is a right angle prism with apex angle of $\alpha = 90°$, while for n=2.4 the maximum difference occurs for $\alpha = 125°$. Although the maximum of curves are in a definite apex angle but the curves are approximately flat to some extend around this specified angles. As a result, there are many different choices available for constituting the plate by adopting different apex angles. Increasing the apex angle behind the angle of maximum difference decreases the transmittance difference, until it totally vanishes for $\alpha = 180°$, as expected for a simple parallelepiped plate.

## IV. Energy transmittance and more realistic conditions

As mentioned previously, there is a different between the transmittance calculated by means of ray tracing method and the transmittance as the ratio of the transmitted energy to the incident energy. It will be more realistic to consider the former as the transmittance of incident photonic flux, regarding each ray as a pseudo-photon propagating along the ray path. Each photon carries a definite energy corresponding to its specified frequency whereas the frequencies are distributed amongst the photons in the form of the Plank's spectral distribution function. The spectral photonic flux density can be derived from the well-known spectral Plank emissive power $E_v(T)$ as

$$\phi_v(T) = \frac{E_v(T)}{hv} = \frac{2\pi v^2}{c^2}\left(e^{\frac{hv}{kT}} - 1\right)^{-1}. \tag{21}$$

The total photonic flux density then reads
$$N(T) = \int_0^\infty \phi_v(T)dv = \sigma'T^3 \tag{22}$$
with $\sigma' = 1.52 \times 10^{15} \frac{Photons}{m^2 sK^3}$, a constant corresponding to the Stefan-Boltzman constant, i.e. $\sigma = 5.67 \times 10^{-8} \frac{J}{m^2 sK^3}$. Equation (22) is counterpart of the Stefan-Boltzman law for the total emissive power of a black-body, i.e.
$$E(T) = \int_0^\infty E_v(T)dv = \sigma T^4. \tag{23}$$

As already mentioned, each ray may be regarded as a representation for a photon in the radiation field. From this point of view the total energy $\epsilon$ corresponding to $n$ photons, each having a specified frequency of $v_i$, equals the sum of individual photon's energies,
$$\epsilon = \sum_{i=1}^{np} hv_i \tag{24}$$
demanded that the frequencies are distributed according to the spectral distribution function (21).

For a non-dispersive medium the transmitted portion of photons also would have the same frequency distribution. Therefore, the remaining problem is the method for giving the desired frequency distribution to the bunch of rays that are transmitted through the prismatic sheet. To this end, we employ the acceptance–rejection method [6] to distribute the number of *n* randomly generated frequencies amongst *n* photons (rays) in the form of distribution function (21). Because of infinitely extended domain of the frequencies we have to truncate the frequencies at a suitable cutoff frequency, which is where the value of photonic fluxes becomes negligible. In the range of common room temperatures we choose $v_c = 3THz$, for which the numerical evaluation of the integral in (22) up to this cutoff frequency gives the same result as one obtains from the Stefan-Boltzman law.

The next step is to test the accuracy of the results from Eq. (24) for the given frequency distribution obtained from numerical acceptance-rejection method. Supposing that the number of *N(T)* thermally radiated photons is equivalent to an amount of *E(T)* radiated energy then from Eqs. (22) and (23) one find for energy of *n* photons

$$\epsilon = \frac{E(T)}{N(T)} n = \frac{\sigma T}{\sigma'} n \qquad (25)$$

A comparison between the energy from the theoretical formula (24) and the corresponding numerical results from Eq. (25) shows the degree of accuracy for our numerical method. With an amount of $n = 10^5$ of photons at $= 300K$, Eq. (25) gives the theoretical value $\epsilon = 1.1191 \times 10^{-15}$ (in arbitrary units) while for the same parameters equation (24) gives the numerical value $\epsilon = 1.1487 \times 10^{-15}$, which is in an acceptable agreement with the theoretical value.

Now we are in a position that we can use our numerical method to estimate the amount of energy transmittance when the ray transmittance is given along the lines of previous section. To get to more realistic conditions it is better to include another important parameter, i.e. the transmitting window for a given material. The transparency range of a material is limited to the frequencies between two thresholds, usually designated by their corresponding wavelengths. For example, the transparency of KBr crystal is in the range of wavelengths from $\lambda_{min} \approx 0.25\mu m$ to $\lambda_{max} \approx 25\mu m$ whereas for ZnSe crystal they range from $\lambda_{min} \approx 1\mu m$ to $\lambda_{max} \approx 18\mu m$. To bring these considerations into account we modify eq. (24) in a new form as

$$\epsilon = \tau\delta \sum_{i=1}^{n} \frac{hc}{\lambda_i} \qquad (26)$$

where $\tau$ is the ray transmittance an $\delta$ is a window function defined as,

$$\delta = \begin{cases} 1, & \lambda_{min} < \lambda_i < \lambda_{max} \\ 0, & else\ where \end{cases}. \qquad (27)$$

Keeping these considerations in mind we calculate the energy transmittance from the ray transmittance values evaluated in previous section. For n=1.5 and choosing the material to be KBr crystal the energy transmittance difference as a result of Eq. (26) is plotted versus the apex angle of prism in Fig. 5. The corresponding quantity for ray transmittance is also plotted for comparison.

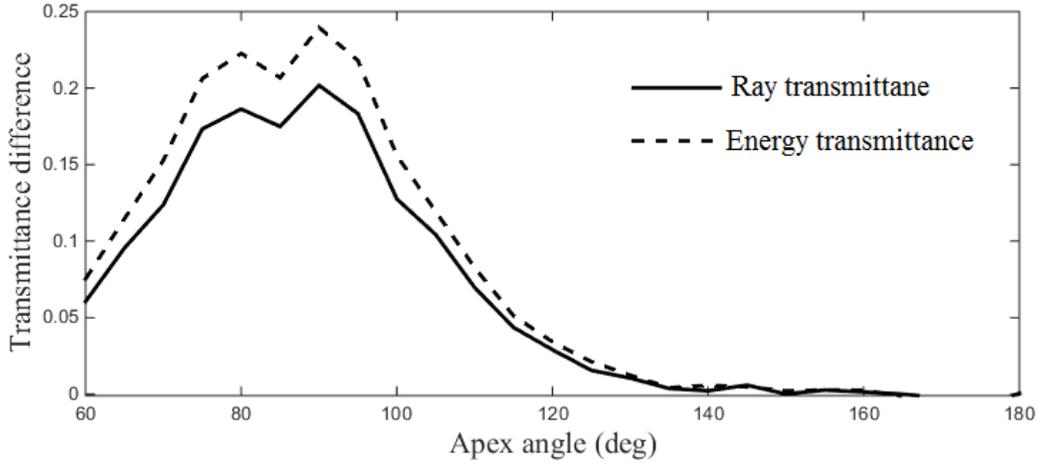

Fig. 5. Graphs of energy and ray transmittance difference vs. apex angle.

As it is inferred from this illustration, the values of transmittance difference for energy exchange are lowered in comparison with the one of ray transferring. Of course the different between the transmittance for two directions still exist and consequently now we can assert certainly that a prismatic sheet is still a worthy sample for the real world bi-transmittance medium in accordance with the thermal radiation energy transfer.

## V. Equilibrium temperature and relaxation time estimation

Another interesting problem is the efficiency of the device when it is used as a radiation rectifier and estimating the temperatures of two sources in thermal equilibrium. A rough estimation for the equilibrium temperature obtains directly from Eq. (12). For the case of KBr crystal with $\tau'_{ba} = 0.40$ and $\tau'_{ba} = 0..58$, when one of source (e.g. the source next to the base) is in contact with a reservoir at room temperature, $T_{eb} = 300K$, the other source (in the absence of the medium emissivity) will achieves an equilibrium temperature of

$$T_{ea} = \left(\frac{\tau'_{ba}}{\tau'_{ab}}\right)^{1/4} T_{eb} \approx 273.4K \;. \tag{28}$$

This is the equilibrium temperature when other heat transfer mechanisms i.e. conduction and convection are neglected.

In practice it is not possible to completely isolate one of the sources such that the radiation heat transfer being the only heat transfer mechanism. Consequently, in a realistic situation there is a competition between the radiation heat transfer, to bring the system into the equilibrium state of Eq. (28) and other mechanism to return back it to the ordinary equilibrium state of $T_{ea} = T_{eb}$. To find the dominant mechanism it is necessary to have an estimation for the time scale of each mechanism. Here we try to find an analytical equation for time evolution of the system. In this perpose, we start with Eq. (6) for the net radiant flux of a source surface. Solving the set of equations (2)-(5) for $J_{m1}$ and $J_{m2}$ and inserting the results in (6) result to the following relation for the rate of net thermal energy intering the surface (1),

$$\frac{dQ}{dt} = A\varepsilon\sigma\tau_{12}(T_1^4 - \delta^4 T_2^4), \tag{29}$$

A being the surface area and $\delta = (\tau_{21}/\tau_{12})^{1/4}$. The absorption of the prismatic sheet is neglected for the sake of simplicity. Taking the source (1) to be a blackened sheet, with $\rho$, d and $c_p$ its mass density, thickness and specific heat of constituting material, respectively, one can set $dQ = \rho A d c_p dT_1$ and then solve the differential equation (29) to get

$$t = \frac{-\rho d c_p}{2\varepsilon\sigma\tau_{12}\delta^3 T_2^3}\left[\tan^{-1}\left(\frac{T_1}{\delta T_2}\right) + \tanh^{-1}\left(\frac{T_1}{\delta T_2}\right) - \tan^{-1}\left(\frac{1}{\delta}\right) - \tanh^{-1}\left(\frac{1}{\delta}\right)\right] \tag{30}$$

This is the required time for the source (1) to reach to the temperature $T_1$, starting from the initial temperature $T_2$, while other source is kept at fixed temperature $T_2$. As an illustrative example, the elapsed time t versus $T_1$ is plotted in Fig. 6 for the case in which the source is taken to be a $d = 2\ mm$ thickness ceramic plate with $\varepsilon = 0.85$, $\rho = 3200\ Kg/m^3$ and $c_p = 710\ J/Kg.K$ and the medium being a KBr prismatic plate. The initial temperature of the surface (1) and the fixed temperature of surface (2) is chosen to be $T_2 = 300\ K$.

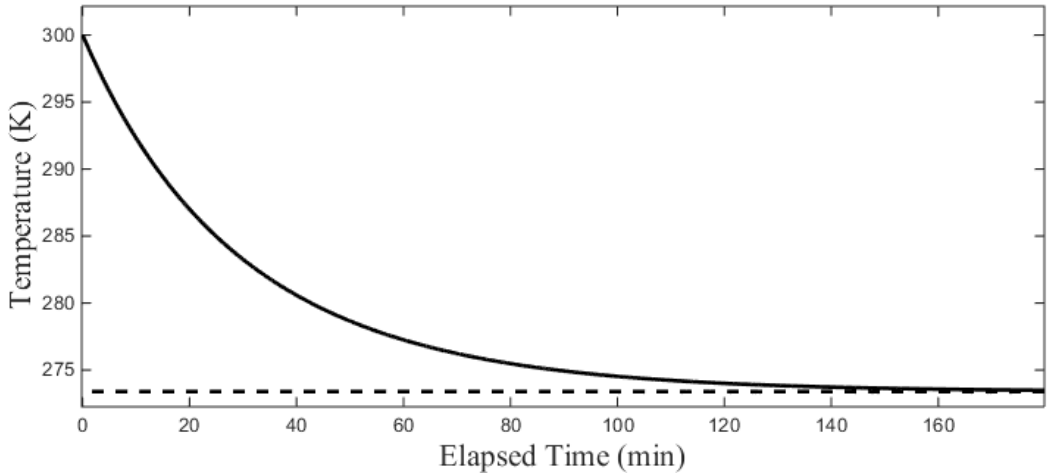

Fig. 5. Variations of an isolated ceramic surface temperature versus time when it is in heat transfer with a reservoir at fixed temperature $T_2 = 300\ K$. The dashed line indicates the final equilibrium temperature ($T_1 \approx 273\ K$).

This illustration shows precisely that such a system never achieves an equilibrium state in a finite time domain, because of the exponential nature of the variation of temperature versus time. Nonetheless, the rate of variation in the beginning of process is considerably fast, such that the half of total temperature difference achieves in almost 20 minutes for the given structure. It shows precisely that the conductive heat transfer is not a serious obstacle and can be overcome by choosing appropriate selection for the restrains that hold the isolated source. The convection heat transfer can also avoided by placing whole of the system in an evacuated chamber.

## VI. Summary and discussion

In spite of the lengthy list of different proposals and plans that have been presented during the last century as the possibility of violation for the second law of thermodynamics in macroscopic scales, still no one has reported an experimental and a real world confirmation for this challenge. Nonetheless, the attraction of the subject, especially because of its relation to the possibility of extracting the internal energy of the world as an endless source of energy, has maintained it as an interesting and alive field for serious scientific investigations and arguments.

In this paper a new challenge with the second law is presented and discussed numerically by means of the bi-transmittance mediums. The theoretical analyzes of the problem in section II confirmed the idea and ended with a relation for the equilibrium temperatures as a function of the medium properties. In section III the prismatic transparent sheet introduced as a medium that has the desired bi-transmittance and its optical properties derived and discussed by means of a numerical simulation method. It showed there that a prismatic transparent sheet has different transmittances for two opposite directions with the values that are dependent on the material and the geometry of the prismatic corrugations on surface of the sheet. Imposing more realistic parameters, such as real energy flow and transmittance window of material, also confirmed again the predicted properties as it discussed in section IV. Finally in section V the applicability of the device proved from the point of view of the relaxation time and the contribution of other competitor phenomena in heat transfer. In summary, the results of our numerical investigations in the context of geometrical optics proved that a prismatic transparent sheet when placed as the medium between two thermally radiating sources works as a radiation rectifier that can establish a temperature gradient across two sources. This is very important and interesting property for a prismatic sheet because its operation is in evident contradiction with the second law of thermodynamics and so deserves to be regarded as a new type of metamaterials. Meanwhile, it is notable that the results are valid only where the geometric optics is valid i.e. where prismatic corrugations are very larger in size than the wavelength of the thermal radiation in use. In contrast, for what are known as microprismatic structures a fully electromagnetic approach is needed to explain their optical properties correctly.

# References

[1] R. Marque´s, F. Marti´n and M Sorolla, *Metamaterials with Negative Parameters*, (John Wiley & Sons, Inc., Hoboken, New Jersey, 2008)

[2] D. F. Wanderwerf , *Applied Prismatic and Reflective Optics*, (SPIE press, Bellingham, Washington USA, 2010).

[3] A. Deinega, I. Valuev, B. Potapkin and Y. Lozovik Y., "Minimizing light reflection from dielectric textured surfaces", Journal of Optical Society of America A, **28**, 5, 2011.

[4] S. C. Yeh et al, "Distribution of Emerged Energy for Daylight Illuminate on Prismatic Elements", Journal of Solar Energy Engineering, **133**, pp 021007-1-9, 2011.

[5] R. Siegel and J. R. Howell, thermal radiation heat transfer (hemisphere publishing corporation, Washington, third edition, 1992)

[6] R. Y. Rubinstein and D. P. Kroese, simulation and the monte carlo method (John Wiley & Sons, Inc., Hoboken, New Jersey, 2017)